\documentclass[aps,superscriptaddress,prd,eprint,nofootinbib,showkeys]{revtex4-2}
\pdfoutput=1
\usepackage{graphicx}
\usepackage[normalem]{ulem} 
\usepackage{dcolumn}
\usepackage{bm}

\usepackage{amsmath,amsfonts,amsthm,amssymb}
\usepackage{graphics}
\usepackage{cancel}
\usepackage{multirow}
\usepackage{longtable}
\usepackage{color}
\usepackage[normalem]{ulem} 

\usepackage{amssymb}

\usepackage{graphicx}
\usepackage{amsmath}
\usepackage{subfigure}
\usepackage{bbold}
\usepackage{yfonts}
\usepackage[colorlinks=true,linktocpage=true,linkcolor=blue,citecolor=blue]{hyperref}
\usepackage{placeins}
\usepackage{bm} 
\usepackage{nicefrac}
\usepackage{slashed}
\usepackage{marginnote}

\begin{document}

\preprint{APS/123-QED}
    
\title{Particle number diffusion in second-order relativistic dissipative hydrodynamics with momentum-dependent relaxation time}

\author{Sunny Kumar Singh}
	\email{sunny.singh@iitgn.ac.in}
	\affiliation{Indian Institute of Technology Gandhinagar,Gandhinagar-382355, Gujarat, India}

\author{Samapan Bhadury}
	\email{samapan.bhadury@uj.edu.pl}
	\affiliation{Institute of Theoretical Physics, Jagiellonian University ul. St. \L ojasiewicza 11, 30-348 Krakow, Poland}        

	\author{Manu Kurian}
	\email{manukurian@iitism.ac.in}
	\affiliation{Department of Physics, Indian Institute of Technology (Indian School of Mines) Dhanbad, Jharkhand 826004, India}

 	\author{Vinod Chandra}
	\email{vchandra@iitgn.ac.in}
	\affiliation{Indian Institute of Technology Gandhinagar,Gandhinagar-382355, Gujarat, India}


\begin{abstract}
 This article explores particle number diffusion in relativistic hydrodynamics using kinetic theory with a modified collision kernel that incorporates the momentum dependence of the particle relaxation time. Starting from the Boltzmann equation within the extended relaxation time approximation (ERTA), we derive second-order evolution equations for the dissipative number current and calculate the associated transport coefficients. The sensitivity of transport coefficients to the particle momentum dependence of the collision time scale of the microscopic interactions in the hot QCD medium is analyzed. For a conformal, number-conserving system, we compare the ERTA-modified transport coefficients for particle diffusion with exact results derived from scalar field theory. With an appropriate parameterization of the relaxation time, we demonstrate the consistency of our analysis and assess the degree of agreement of the results with the exact solutions from scalar field theory. The relaxation times for the shear and number diffusion evolution equations are seen to be distinct in general when the momentum dependence of the relaxation time is taken into consideration.
\end{abstract}

\maketitle

\section{Introduction}
Energetic heavy-ion collision experiments conducted at the Relativistic Heavy-Ion Collider (RHIC) and Large Hadron Collider (LHC) create a strongly interacting nuclear matter-the quark-gluon plasma (QGP), under extreme conditions of energy density and temperature. The success of theoretical models based on relativistic hydrodynamics in describing the evolution of the QGP is widely regarded as one of the remarkable achievements in this field~\cite{ Heinz:2013th, Gale:2013da,Jaiswal:2016hex,Florkowski:2017olj}. These models can effectively capture the particle spectra, collective flow, etc., and provide insights into the properties of the hot nuclear matter.  Consequently, substantial research has been devoted to the fluid-like behavior of the QGP, particularly through extraction of its transport properties~\cite{Teaney:2003kp,Romatschke:2007mq,Noronha-Hostler:2013gga,Ryu:2015vwa,Heffernan:2023gye, Wagner:2024fry, Palni:2024wdy}.

The evolution of a relativistic fluid (for example, the QGP evolution in heavy-ion collisions) that is not too far from its local equilibrium state, can be described through relativistic hydrodynamics which can be formulated as an expansion in gradients of the primary fluid-dynamical variables. The Knudsen number serves as the expansion parameter of the underlying hydrodynamic theory for the QGP evolution. The zeroth-order expansion corresponds to a non-dissipative fluid in which the fluid is assumed to be in the local thermodynamic equilibrium, i.e, an ideal fluid. The first-order theory-the relativistic Navier-Stokes theory, takes into account the dissipative terms of first order in Knudsen number. However, the Navier-Stokes theory is found to be unstable and acausal~\cite{AIHPA_1965__2_1_21_0,Hiscock:1985zz,Denicol:2008ha}, making it unsuitable for a hydrodynamical description of the QGP. To overcome these challenges, the second-order M\"{u}ller-Israel-Stewart theory was introduced~\cite{Muller:1967zza,Israel:1976tn,Israel:1979wp}. Moreover, some recent advances, generally referred to as Bemfica, Disconzi, Noronha, and
Kovtun (BDNK) theory, have demonstrated that with appropriate modifications of the energy-momentum tensor, first-order theory can be made causal and stable~\cite{Bemfica:2019knx,Hoult:2020eho}. However, its phenomenological implications remain to be fully developed \cite{Pandya:2021ief}. The applicability of the dissipative hydrodynamic framework is inherently tied to the assumptions and approximations of the underlying microscopic theory from which it is derived. The second-order theory in the original M\"{u}ller-Israel-Stewart work was derived by analyzing entropy production. In more recent approaches, hydrodynamic equations are obtained by starting from the Boltzmann equation, utilizing methods such as the Chapman-Enskog expansion~\cite{Jaiswal:2013npa,Panda:2020zhr} or the method of moments~\cite{ Denicol:2010xn,Denicol:2012cn}. 

Dissipative quantities play a crucial role in describing the non-equilibrium dynamics of the QGP. Extensive research has focused on understanding the evolution of the shear stress tensor and bulk viscous pressure in the hot QCD medium. In contrast, much less attention has been paid to the role of dissipative charge currents. This disparity arises largely from the assumption that net-baryon density is negligible at very high collision energies. However, this assumption breaks down for low-energy collisions, such as those explored in the Beam Energy Scan (BES) program at the RHIC or the upcoming experiments at the Facility for Antiproton and Ion Research (FAIR). At these energies, the net-baryon density becomes significant, and the effects of dissipative charge currents can no longer be ignored. Charge diffusion may play a critical role in the hydrodynamic evolution of the QCD medium, making its inclusion in theoretical models essential to accurately describe the system's behavior. In recent years, efforts have been made in this area, with studies addressing the formulation of hydrodynamic theory at finite chemical potential~\cite{Denicol:2010xn,Jaiswal:2015mxa,Fotakis:2019nbq,Dash:2022xkz,Daher:2024vxk}, as well as numerical simulations exploring the influence of net baryon diffusion on observables such as particle spectra and flow coefficients~\cite{Denicol:2018wdp,Du:2019obx,An:2021wof}.

One of the crucial steps in deriving the hydrodynamic equations from kinetic theory is to address the complexities of effectively incorporating microscopic interactions in the medium. The Anderson-Witting Relaxation Time Approximation (RTA)~\cite{Anderson:1974nyl} has emerged as a widely used framework that characterizes the collision kernel of the Boltzmann equation with the relaxation time parameter.  The RTA has been extensively utilized in the formulation of relativistic hydrodynamics, and in the computation of transport coefficients~\cite{ Jaiswal:2014isa, Denicol:2014xca, Blaizot:2017lht, Blaizot:2017ucy, Kurian:2018qwb, Bhadury:2020ivo,Rath:2020wgj,Gowthama:2020ghl,Ghosh:2022xtv,Dey:2020awu,Ghosh:2022vjp, Weickgenannt:2024ibf}.  Some of the initial studies on the diverging nature of hydrodynamic gradient expansion~\cite{Heller:2013fn, Buchel:2016cbj} were carried out using the RTA \cite{Heller:2013fn}. Moreover, the attractor solution in hydrodynamics~\cite{Jaiswal:2021uvv, Jaiswal:2022udf, Vyas:2022hkm} and the exact numerical solutions~\cite{Denicol:2014xca, Florkowski:2014sfa, Chattopadhyay:2018apf} to the relativistic Boltzmann equation were also found using the RTA collision kernel. The standard implementation of the RTA typically assumes that the relaxation time is independent of the particle's momentum. However, this assumption is not valid for realistic systems, where the collision timescale depends on the specifics of microscopic interactions. Recent studies have introduced significant improvements to the conventional RTA, addressing these limitations and ensuring compatibility of the relativistic Boltzmann equation with the conservation laws
~\cite{Rocha:2021zcw, Mitra:2020gdk,Rocha:2022fqz,Shaikh:2024tjn}. Refs.~\cite{Teaney:2013gca,Dash:2021ibx, Dash:2023ppc} introduced the Extended Relaxation Time Approximation (ERTA), which incorporates the momentum-dependent relaxation time and provides a systematic way of deriving hydrodynamic equations. Notably, the viscous hydrodynamical evolution and the associated transport coefficients are observed to be sensitive to the momentum dependence of the relaxation time~\cite{Dash:2023ppc,Singh:2024leo,Bhadury:2024ckc, Kamata:2022ola}. The precise functional relationship between relaxation time and particle momentum is intricate and challenging to determine. In Ref.~\cite{Dusling:2009df}, the authors discussed examples of several theories where the relaxation time can depend on the particle momentum. While most of the theories may lie between two limiting cases where the relaxation time is either constant (called \textit{linear ansatz}) or proportional to momentum linearly (called \textit{quadratic ansatz}), there may exist theories where this relationship is non-trivial. In the case of linear ansatz, the energy loss of high-momentum particles increases linearly with momentum. In contrast, in the case of quadratic ansatz, the energy loss of high-momentum particles approaches a constant. To incorporate the momentum dependence of the relaxation time into our analysis, we adopted a general parameterized form, focusing on the regime relevant to the QCD medium.

In this paper, we investigate particle number diffusion within relativistic second-order dissipative hydrodynamics, incorporating a momentum-dependent thermal relaxation time. Using the ERTA framework, we derive the second-order evolution of the dissipative particle number current for the first time. With this modified collision kernel, we compute the transport coefficients associated with number current transport in a general reference frame, incorporating the landau matching conditions. Exact analytical solutions to the Boltzmann equation remain challenging to obtain.
However, a recent study~\cite{Denicol:2022bsq,Rocha:2023hts} provides the exact transport coefficients for a massless self interacting $\lambda \phi^4$ theory. For a conformal, number conserved system, we perform a direct comparison between the ERTA transport coefficients for particle diffusion and the exact scalar field theory results, with an appropriate choice of a parameterized-relaxation time.     

The paper is organized as follows. In Section~\ref{Section1}, we derive the evolution of particle number diffusion for a number conserving  conformal system using kinetic theory with the modified collision kernel within the ERTA framework. In Section~\ref{Section2}, we examine the particle diffusion relaxation time in the current framework and its significance. The latter part of this section is dedicated to comparing our results with the exact solutions from scalar field theory. We summarize the analysis in Section~\ref{Section3}. The definitions of various thermodynamic integrals and their useful relations are presented in Appendix~\ref{Appendix1}.
\\
{\it Notation and Conventions}:  Throughout, we  follow natural units $c=k_B=\hbar=1$ and the metric tensor is defined as $g_{\mu\nu}={\text {diag}}\,(1,-1,-1,-1)$. The four-velocity of the fluid element is represented by $u_\mu$ with the normalization condition $u^\mu u_\mu=1$. In the local rest frame of the fluid, we have $u^{\mu}=(1,0,0,0)$. We define the two-rank projection operator as $\Delta^{\mu\nu} = g^{\mu\nu} - u^\mu u^\nu$ with $\Delta^{\mu\nu} u_\mu=0$ and  rank-four projection operator perpendicular to $u^\alpha$ and $\Delta^{\alpha\beta}$ as $\Delta^{\mu\nu}_{\alpha\beta}\equiv\frac{1}{2}(\Delta^\mu_\alpha\Delta^\nu_\beta +\Delta^\mu_\beta\Delta^\nu_\alpha)-\frac{1}{3}\Delta^{\mu\nu}\Delta_{\alpha\beta}$.

\section{ Charge diffusion in dissipative hydrodynamics}\label{Section1}
In the present analysis, we consider a system of massless particles with a finite chemical potential$-$a conformal, particle number conserving system$-$ where the constituent particles follow Maxwell-Boltzmann statistics. The hydrodynamic evolution of such a system is determined by the conservation equations of the net particle number current along with the energy and momentum. The particle number current and energy-momentum tensor of the system can be expressed in terms of the single-particle phase-space distribution $f(x,p)$ as \cite{groot1980relativistic, cercignani2002relativistic},
\begin{align}
    &N^\mu = \int \mathrm{dP} p^\mu f, \label{N^m,T^mn-fromf}
    && T^{\mu\nu} = \int \mathrm{dP} p^\mu p^\nu f,
\end{align}
where the invariant phase-space integral measure is given by, $dP = d_g\frac{ d^3 |\textbf{p}|}{\left(2\pi\right)^3 p^0}$ with $d_g$ being the degeneracy factor. projections of the number four-current along the
fluid velocity and orthogonal to it, one can decompose $N^\mu$ as,
\begin{align}
    N^\mu &= n u^\mu + n^\mu, \label{N^m-decomp}
\end{align}
 Where $u^\mu$ is the four-velocity of a fluid element defined in the Landau frame satisfying the condition, $u_\nu T^{\mu\nu}=\epsilon u^\nu$, with $\epsilon$ and $n$ being the energy density and net particle number density respectively in the local rest frame of the fluid,  and $n^\mu$ is the dissipative term that denotes the number diffusion current. Similarly, the  energy-momentum tensor can be decomposed as, 
\begin{align}
    T^{\mu\nu} = \epsilon u^\mu u^\nu - P \Delta^{\mu\nu} + \pi^{\mu\nu}, \label{T^mn-decomp}
\end{align}
 with $P$ and $\pi^{\mu\nu}$ being the thermodynamic pressure and shear viscous stress respectively. 
The hydrodynamic equations for the evolution of $\epsilon$, $n$, and $u^\mu$ are obtained by taking the projections of the energy-momentum conservation, $\partial_\mu T^{\mu\nu}=0$ and number current conservation, $\partial_\mu N^\mu=0$ along and perpendicular to the direction of fluid flow,
\begin{align}
    \dot{\epsilon} + (\epsilon +  P)\theta - \pi^{\mu\nu}\sigma_{\mu\nu} &= 0,\label{Heqs1}\\
    (\epsilon+ P)\dot{u}^\mu - \nabla^\mu P + \Delta^\mu_\nu \partial_\gamma \pi^{\nu\gamma} &= 0,\label{HeqsM}\\
    \dot{n} + n\theta + \partial_\mu n^\mu = 0, \label{Heqs}
\end{align}
where $\dot{A}=u^\mu\partial_\mu A$ is the co-moving derivative of $A$ which becomes the purely temporal derivative in the local rest frame of the fluid, and $\nabla^\mu = \Delta^{\mu\nu}\partial_\nu$ is the spacelike derivative which becomes the purely spatial derivative in the local rest frame of the fluid.  In Eq.~(\ref{Heqs1}), $\sigma^{\mu\nu}=\Delta^{\mu\nu}_{\alpha\beta} \nabla^\alpha u^\beta$ is the velocity stress tensor and $\theta=\partial_\mu u^\mu$ is the expansion scalar. For a system of massless particles following the Maxwell-Boltzmann distribution in the local rest frame of the fluid at equilibrium\footnote{For anti-particles we have to change $\mu\to - \mu$. But in the present case, we do not consider anti-particles, as we demand the particle number to be conserved.}, the equilibrium distribution function for the particles is given by $f_0 =e^{-\beta(u \cdot p)+\alpha}$, where $\beta = 1/T$ and $\alpha = \mu/T$ are the inverse of temperature $T$ and ratio of chemical potential, $\mu$ to the temperature, $T$ respectively. The equilibrium number density, energy density and pressure are then given by:
\begin{align} 
    \epsilon_0 &= u_\nu u_\mu \int \mathrm{dP} p^\mu p^\nu f_0 = \frac{3 d_g e^\alpha}{\pi^2 \beta^4} \label{e1}\\
    n_0 &= u_\nu \int \mathrm{dP} p^\nu f_0 = \frac{d_g e^\alpha}{\pi^2 \beta^3} \\
    P &= - \frac{1}{3}\Delta_{\mu\nu} \int \mathrm{dP} p^\mu p^\nu f_0 = \frac{d_g e^\alpha}{\pi^2 \beta^4}. \label{e,n,P-def}
\end{align}

The equation of state for this system of massless particles is given by $P=\epsilon/3$. Using the Landau matching conditions given by, $\epsilon=\epsilon_0$ and $n=n_0$,
the expressions for the spacelike and co-moving derivatives of $\alpha$ and $\beta$, can be obtained by substituting Eqs.~(\ref{e1})-(\ref{e,n,P-def}) into Eqs.~(\ref{Heqs1})-(\ref{Heqs}) as,
\begin{align}
    \dot{\alpha} &= -A_n \partial_\mu n^\mu - A_\Pi \pi^{\mu\nu}\sigma_{\mu\nu},\\
    \dot{\beta} &= \frac{\beta\theta}{3} - D_n \partial_\mu n^\mu - D_\Pi \pi^{\mu\nu}\sigma_{\mu\nu},\\
    \nabla^\mu \beta &= -\beta \dot{u}^\mu + \chi\nabla^\mu \alpha - \xi \Delta^\mu_\nu \partial_\gamma \pi^{\nu\gamma}, \label{Da,Db,nabla-b:def}
\end{align}
where, the coefficients are defined as,
\begin{align}
    &A_n = \frac{4}{n}, && A_\Pi =\frac{\beta}{n},  &&& D_n = \frac{\beta}{n}, \nonumber\\
     &D_\Pi = \frac{\beta}{\epsilon}, 
    &&\chi = \frac{n}{\epsilon+P} = \frac{\beta}{4}, &&&\xi = \frac{\beta}{\epsilon+P} = \frac{3\beta}{4\epsilon}. \label{Coeffs.-def}
\end{align}
These relations will be used while deriving the evolution equation for $n^\mu$ in the next section.  In a prior study~\cite{Dash:2023ppc}, the second-order evolution equation for the shear stress tensor, incorporating the effects of a modified collision kernel, has been extensively studied. In the present work, we extend this analysis to the evolution of the number current, which is essential for an accurate description of the hydrodynamic evolution of the QCD medium at finite chemical potential. To achieve this, we consider a system near local equilibrium where the distribution function can be expressed as $f=f_0+\delta f$ where the deviation from the equilibrium distribution, $\delta f$ satisfies the condition $|\delta f|\ll f_0$. The shear stress tensor and number diffusion current can be further defined as,
\begin{align}
    \pi^{\mu\nu} &= \Delta^{\mu\nu}_{\rho\gamma}\int \mathrm{dP} p^\rho p^\gamma \delta f. \label{pi^nm} \\
    n^\mu &= \Delta^\mu_\nu \int \mathrm{dP} p^\nu \delta f. \label{pi^nm,n^nm-Def}
\end{align}
The first step in the derivation of the number current evolution is the estimation of the appropriate non-equilibrium part of the distribution function $\delta f$. 

\subsection{Non-equilibrium distribution function within ERTA at finite $\mu$}
The non-equilibrium correction $\delta f$ can be obtained by solving the relativistic Boltzmann equation. The evolution of the phase-space distribution function is described by the Boltzmann equation as,
\begin{align}
    p^\mu \partial_\mu f= C[f], \label{BoltzmanEq}
\end{align}
where $C[f]$ is the collision kernel that encodes all the interactions between the colliding particles in the system and can be defined as,
\begin{align}
    C[f] = \frac{1}{2} \int dP' \; dK \; dK' 
    \; W_{pp'\rightarrow kk'}(f_k f_k' - f_p f_p'). \label{collisionKernel}
\end{align}
The above expression is valid for $2\longleftrightarrow 2$ elastic collisions where $W_{pp'\rightarrow kk'}$ is the collision transition rate. Here, the subscripts on the distribution function and the transition rate indicate the particle momenta. The Boltzmann equation with the above collision kernel becomes an integro-differential equation which becomes difficult to solve for almost all kinds of interactions. To address this, approximations are used to simplify the collision term in the linearized regime. One notable approach is the Relaxation Time Approximation (RTA), introduced by Anderson and Witting \cite{Anderson:1974nyl}. With the assumption that the collisions in the system restore the local distribution function to its local equilibrium value exponentially, we have
\begin{align}
    C[f]= -\frac{u\cdot p}{\tau_R(x)} (f- f_0), \label{RTA-def}
\end{align}
with $\tau_R$ as the thermal relaxation time parameter. The conventional RTA
assumes $\tau_R(x)$ to be independent of particle momentum and leads to the conservation of number current and energy-momentum tensor by taking the zeroth and first moments of the collision term respectively only when the Landau matching conditions are employed.  In contrast, the timescale of collisions depends on the microscopic interactions within the medium, which can introduce a momentum-dependent relaxation time $\tau_R(x, p)$. 
However, incorporating $\tau_R(x, p)$ in the conventional RTA makes it incompatible with the conservation laws, even in the Landau frame. This sets the motivation for a modified collision kernel that can incorporate a momentum-dependent relaxation time while maintaining the conservation laws in the Landau frame. In the present analysis, we employ the recently developed Extended Relaxation Time Approximation (ERTA)~\cite{Dash:2021ibx, Dash:2023ppc} that takes into account the particle energy dependent-relaxation time.
The Boltzmann equation with ERTA collision term becomes,
\begin{align}
    p^\mu\partial_\mu f = -\frac{u\cdot p}{\tau_R (x,p)}(f-f_0^*), \label{ERTA-BoltzmanEq}
\end{align}
where $f_0^*=e^{-\beta^*(u^*\cdot p)+\alpha^*}$ is the local equilibrium distribution in the thermodynamic frame \cite{Dash:2021ibx}. Here, $u^*_\mu$, $T^*$, and $\mu^*$ are the fluid four-velocity, temperature, and chemical potential respectively in this thermodynamic frame. The interpretation of the thermodynamic frame as well as the distinction between this and the hydrodynamic frame is explained in detail in Ref.~\cite{Dash:2021ibx}. Using a Chapman-Enskog-like gradient expansion method, we can expand the non-equilibrium correction to the local distribution function in hydrodynamic equilibrium in various orders in gradients as,
\begin{align}
    \delta f= \delta f_{(1)}+\delta f_{(2)}+ \cdots , \label{del-F-expand}
\end{align}
Where $\delta f_{(i)}$ is the term consisting of $i^{th}$ order gradients of the primary hydrodynamic variables, $\alpha$, $\beta$ and $u^\mu$. Each derivative of these hydrodynamic variables is proportional to the inverse of the macroscopic length scale, $L$ over which these primary fluid-dynamical variables varies. With the microscopic length scale $\lambda^i$ contained in the $i^{th}$ term, the $\delta f_{(i)}$ terms are proportional to powers of the Knudsen number, $(Kn)^i \sim (\lambda / L)^i$. Hence, assuming that the microscopic length scales are much smaller than the macroscopic length scales in the system, i.e, $Kn << 1$, the gradient expansion thus considered is equivalent to the power counting scheme in terms of the Knudsen number, $Kn \propto \partial$ which can be used to truncate the above expansion of $\delta f$ at any desired order\footnote{From here on, we denote any such $i^{th}$ order derivative term with the notation, $\mathcal{O}(\partial^i)$.}. With $\delta f^*=f^*_0-f_0$, the ERTA-modified Boltzmann equation leads to the following form of $\delta f$,
\begin{align}
    \delta f= -\frac{\tau_R(x,p)}{u \cdot p} p^\mu \partial_\mu f + \delta f^*. \label{delF-general}
\end{align}
The $\delta f^*$ acts as a counter-term to ensure the conservation laws and is determined by imposing the Landau frame and the Landau matching conditions. The form of distribution function up to the second order with the vanishing chemical potential is presented in Ref.~\cite{Dash:2023ppc}. Since the present focus is on a charge-conserved system,  it is necessary to analyze the matching conditions and estimate the distribution function up to second-order at a finite $\mu$.  We begin with the evaluation of the first-order to distribution function, $\delta f_{(1)}$. Up to first-order in gradients, Eq.~(\ref{delF-general}) can be written as,
\begin{align}
    \delta f_{(1)}= -\frac{\tau_R(x,p)}{u \cdot p} p^\mu \partial_\mu f_0 + \delta f^*_{(1)}. \label{delF-1stOrder}
\end{align}
With $T^*=T+\delta T$, $ \mu^*=\mu+\delta \mu$ and $ u^*_\mu=u_\mu+\delta u_\mu $, we can Taylor expand $f_0^*(u_\mu^*,T^*,\mu^*)$ around  $T$, $\mu$ and $u_\mu$ to obtain the following expression for $\delta f_{(1)}^*$,
\begin{align}
  &\delta f^*_{(1)} = \left[ -\frac{(\delta u \cdot p)}{T} + \frac{(u \cdot p - \mu)}{T^2} \delta T + \frac{\delta \mu}{T} + \mathcal{O}(\partial^2)+ \cdots \right] f_0.  \label{delF_1*-TaylorExpansion}
\end{align}
By employing Eq.~(\ref{delF_1*-TaylorExpansion}) and  Eq.~(\ref{delF-1stOrder}), the first-order distribution can be written as,
\begin{align}
    &\delta f_{(1)} = \tau_R \left[\left(\frac{n}{\epsilon+P} - \frac{1}{u \cdot p}\right) p_\mu \left(\nabla^\mu \alpha\right) + \frac{\beta p^\mu p^\alpha \sigma_{\mu \alpha}}{(u \cdot p)} \right] f_0 + \left[ -\frac{(\delta u \cdot p)}{T} + \frac{(u \cdot p - \mu)}{T^2} \delta T + \frac{\delta \mu}{T}\right] f_0,\label{delF_1-FullExpr}
\end{align}    
where $\delta T$, $\delta \mu$, and $\delta u^\mu$, can be determined by imposing the matching conditions, $\epsilon=\epsilon_0$ and $n=n_0$ along with the Landau frame condition $u_\nu T^{\mu\nu}=\epsilon u^\mu$. Then up to first-order in spacetime gradient, we can obtain the following relations~\cite{Dash:2021ibx},
\begin{align}
  &\delta u^\mu=C\beta \left(\nabla^\mu \alpha\right) + \mathcal{O} (\partial^2),
  &&
  \delta \mu=\mathcal{O} (\partial^2),
 &&&
  \delta T= \mathcal{O} (\partial^2). \label{1stOrdMatch}
\end{align}
Here, the form of $C$ can be expressed in terms of the thermodynamic integrals as,
\begin{align}
    C=\frac{1}{\beta^2 I_{31}}\left( \frac{n}{\epsilon+P}K_{31} - K_{21} \right). \label{C-Def}
\end{align}
The definition of the thermodynamic integrals is presented in Appendix~\ref{Appendix1}. Using the above form of $\delta u^\mu$, $\delta \mu$ and $\delta T$ up to first order in gradients, the first order shear viscosity, $\pi^{\mu\nu}_{(1)}$ and number diffusion current, $n^\mu_{(1)}$ were determined using the definitions given in Eq.~(\ref{pi^nm}) and Eq.~(\ref{pi^nm,n^nm-Def}) as \cite{Dash:2021ibx}:
\begin{align} \label{eta}
    \pi^{\mu\nu}_{(1)} = 2 \eta \sigma^{\mu \nu} , \quad \quad \eta= \frac{2 K_{32}}{T}.
\end{align}
\begin{align} \label{kappa}
    n^{\mu}_{(1)} = \kappa \nabla^\mu \alpha, \quad \quad \kappa= -C\beta^2 I_{21} + \frac{n}{\epsilon+P} K_{21} - K_{11}.
\end{align}

Where $\eta$ and $\kappa$ are the first order shear and diffusion coefficients respectively. With $\delta f_{(2)}$ being the the non-equilibrium correction to the local distribution containing only second order terms in gradients, the total correction till second-order $\Delta f_{(2)}=\delta f_{(1)}+\delta f_{(2)}$, can be further evaluated by keeping terms up to second order in gradients,
\begin{align}
    \Delta f_{(2)} = -\frac{\tau_R(x,p)}{u\cdot p}p^\mu \left(\partial_\mu f_0\right) -\frac{\tau_R(x,p)}{u\cdot p}p^\mu \left(\partial_\mu \delta f_{(1)}\right) + \Delta f^*_{(2)},\label{delF-2ndOrder}
\end{align}
where,
\begin{align}
    \Delta f^*_{(2)} =  \left[-\frac{\Delta u \cdot p}{T} +\left(\frac{u \cdot p - \mu}{T^2}\right) \Delta T + \frac{\Delta \mu}{T} + \frac{(\Delta u \cdot p)^2}{2T^2} + \mathcal{O}(\partial^3) \right] f_0. \label{f2-TaylorExp}
\end{align}
Here, $\Delta u_\mu$, $\Delta T$ and $\Delta \mu$ contains both first and second order contributions. From the first order matching conditions, we know that $\Delta T$ and $\Delta \mu$ are at least second order in gradients. Therefore, higher-order terms are not considered here, as any cross-term contribution would be at least $\mathcal{O}(\partial^3)$. On the other hand, $\Delta u_\mu$ has a first-order contribution, which allows for the inclusion of the term, $\left(\Delta u\cdot p\right)^2$ in Eq.~(\ref{f2-TaylorExp}), making it of the order $\mathcal{O}(\partial^2)$ or higher. Using Eq.~(\ref{f2-TaylorExp}) and by expanding other terms of Eq.~(\ref{delF-2ndOrder}), we obtain the non-equilibrium part of the distribution function up to second-order at a finite $\mu$ as follows,

\begin{align}
    \Delta & f_{(2)} = \Bigg\{-\frac{\Delta u \cdot p}{T} +\frac{(u \cdot p - \mu)}{T^2} \Delta T + \frac{\Delta \mu}{T}+ \frac{(\Delta u \cdot p)^2}{2T^2}\Bigg\}f_0 
    +\tau_R \Bigg[\left(\frac{n}{\epsilon+P} - \frac{1}{(u \cdot p)}\right) p^\rho \nabla_\rho \alpha + \frac{\beta p^\rho p^\gamma \sigma_{\rho \gamma}}{(u \cdot p)} \nonumber\\
    &-\frac{\beta}{\epsilon+P} \Big\{p^\rho \nabla^\gamma \pi_{\rho\gamma} - p^\rho \pi_{\rho\gamma} \dot{u}^\gamma \Big\}+\Big(A_n -D_n(u\cdot p)\Big)\partial_\rho n^\rho +\left\{A_\Pi -\left(D_\Pi+\frac{\beta}{\epsilon+P}\right) (u \cdot p)\right\} \pi^{\rho\gamma}\sigma_{\rho\gamma} \Bigg] f_0 \nonumber\\
    &-\frac{\tau_R}{T}\Bigg[ \dot{\tau}_R\frac{p^\rho p^\gamma}{u \cdot p} \sigma_{\rho\gamma} + (\nabla_\alpha \tau_R) \frac{p^\rho p^\gamma p^\lambda}{(u \cdot p)^2}\sigma_{\gamma\lambda} +C\beta \left( \frac{n}{\epsilon+P} - \frac{1}{u \cdot p} \right)p^\rho p^\gamma \nabla_\rho \alpha \nabla_\gamma \alpha + C\beta^2 \frac{p^\rho p^\gamma p^\lambda}{u \cdot p} \sigma_{\rho\gamma} \nabla_\lambda \alpha \nonumber\\
    &-C\beta p^\rho D(\nabla_\rho \alpha) 
     - C\beta \frac{p^\rho p^\gamma}{u \cdot p} \nabla_\rho \nabla_\gamma \alpha -T p^\rho \nabla_\rho \alpha D(C\beta^2) + T \dot{\tau}_R \left(\frac{n}{\epsilon+P} - \frac{1}{u \cdot p} \right) p^\rho \nabla_\rho \alpha - T \frac{p^\rho p^\gamma}{u \cdot p} (\nabla_\rho \alpha) \nabla_\gamma (C \beta^2)  \nonumber\\
    &+T\left(\frac{n}{\epsilon+P}-\frac{1}{u \cdot p}\right) \frac{p^\rho p^\gamma}{u \cdot p} (\nabla_\rho \alpha)(\nabla_\gamma \tau_R)\Bigg]f_0 -\frac{\tau_R^2}{T} \Bigg[ \frac{2\theta}{3} \frac{p^\rho p^\gamma}{u \cdot p} \sigma_{\rho\gamma} + \frac{p^\rho p^\gamma}{u \cdot p} \dot{\sigma}_{\rho\gamma} + \frac{ p^\rho p^\gamma p^\lambda}{(u \cdot p)^2} (\nabla_\lambda \sigma_{\rho\gamma}) \nonumber\\
    &- \frac{2  p^\rho p^\gamma p^\lambda}{(u \cdot p)^2} \sigma_{\rho \gamma} \dot{u}_\lambda - \left(\frac{1}{T} + \frac{1}{u \cdot p}\right) \left( \frac{p^\rho p^\gamma \sigma_{\rho \gamma}}{u\cdot p} \right)^2
    -2\left(\frac{n}{\epsilon+P} - \frac{1}{u \cdot p}\right) \frac{p^\rho p^\gamma p^\lambda}{u \cdot p} \sigma_{\rho \gamma} \nabla_\lambda \alpha + T\left(\frac{n}{\epsilon+P}\right) \frac{ p^\rho p^\gamma p^\lambda}{(u \cdot p)^2} \sigma_{\rho \gamma} \nabla_\lambda \alpha \nonumber\\
    &+ T \frac{p^\rho p^\gamma p^\lambda}{(u \cdot p)^3} \sigma_{\rho \gamma} \nabla_\lambda \alpha - \frac{\theta T}{3(u \cdot p)} p^\rho \nabla_\rho \alpha -T\left(\frac{n}{\epsilon+P} - \frac{1}{u \cdot p}\right)^2 p^\rho p^\gamma \nabla_\rho \alpha  \nabla_\gamma \alpha +T\left(\frac{n}{\epsilon+P} - \frac{1}{u \cdot p}\right) p^\rho D(\nabla_\rho \alpha) \nonumber\\
    &+T \left(\frac{n}{\epsilon+P} - \frac{1}{u \cdot p}\right) \frac{p^\rho p^\gamma}{u \cdot p}\nabla_\rho \nabla_\gamma \alpha  + T D\left(\frac{n}{\epsilon+P}\right) p^\rho \nabla_\rho \alpha +T \frac{p^\rho p^\gamma}{(u \cdot p)^2} \dot{u}_\rho \nabla_\gamma \alpha +T \frac{p^\rho p^\gamma}{u \cdot p}     \nabla_\rho \left(\frac{n}{\epsilon+P}\right) \nabla_\gamma \alpha \Bigg] f_0. \label{Full-DelF2}
\end{align}
This expression of $\Delta f_{(2)}$ can be utilized to derive the evolution equation for number diffusion current, $n^\mu$. However, the above expression of $\Delta f_{(2)}$ contains the unknown quantities, $\Delta u^\mu$, $\Delta T$, and $\Delta \mu$. Thus, our next task will be to determine these quantities with the help of the Landau frame and matching conditions. It should be noted that the derivation of the evolution equation of $n^\mu$, up to second-order in gradient, does not require $\Delta T$ and $\Delta\mu$, in the following we will confine our discussion to $\Delta u^\mu$ only. 

\subsection{Matching conditions and evolution equation for diffusion current}

In a prior study~\cite{Singh:2024leo}, we derived the evolution equation for the shear stress tensor by incorporating a momentum-dependent relaxation time. In that case, the explicit evaluation of $\Delta f_{(2)}^*$
was not required, as the contribution of the counter terms vanished at second-order for the shear tensor evolution up to second-order in gradient. However, in the present study, it is necessary to determine the form of $\Delta f_{(2)}^*$ as $n^\mu$ will get contribution from $\Delta u^\mu$. By employing Eq.~(\ref{delF-2ndOrder}) in Eq.~(\ref{pi^nm,n^nm-Def}), we can express $ n_\mu$ as follows,  
\begin{align}
    n_\mu &= \Delta_{\mu\nu}\int \mathrm{dP} p^\nu \Delta f_{(2)} = n^*_\mu + \bar{n}_\mu. \label{n^m-def}
\end{align}
Here, $ n^*_\mu$ is the contribution from $\Delta f_{(2)}^*$ part that arises due to the counter terms and $\bar{n}_\mu$ is the contribution from the remaining terms of Eq.~(\ref{delF-2ndOrder}) to the number diffusion current and are defined as,
\begin{align}
     & n^*_\mu = \Delta_{\mu\nu}\int \mathrm{dP} p^\nu \Delta f^*_{(2)},\label{form1}\\
    & \bar{n}^\mu = \Delta^\mu_\nu \int \mathrm{dP} p^\nu \Bigg[ -\frac{\tau_R}{u \cdot p}p^\rho \partial_\rho f_0- \frac{\tau_R}{u\cdot p} p^\rho \partial_\rho (\delta f_{(1)})\Bigg].\label{form2}
\end{align}
By substituting the form of $\Delta f^*_{(2)}$ as defined in Eq.~(\ref{f2-TaylorExp}) into Eq.~(\ref{form1}), we obtain the contribution from counter terms as,   
\begin{align}
    n_*^\mu =-\frac{1}{T} I_{21} \Delta u^\mu + \frac{1}{T} I_{21} (u \cdot \Delta u) u^\mu. \label{n^m*}
\end{align}
Notably, $\Delta T$ and $\Delta \mu$ do not contribute to number diffusion evolution. 
Here, we use the fact that $u \cdot \Delta u \sim \mathcal{O}(\partial^2)$ since $u^*\cdot u^* =1$ and we ignored all the terms of order $\mathcal{O}(\partial^3)$ and higher. 
The form of $\Delta u^\mu$ can be evaluated from the Landau frame condition and the energy matching condition. 
The Landau frame condition, $u_\nu T^{\mu\nu} = \epsilon u^\mu$ leads to the following constraint,
\begin{align}
    u_\nu \int \mathrm{dP} p^\mu p^\nu \Delta f_{(2)} = 0. \label{LandauFrame-Cond}
\end{align}
Substituting the exact form of $\Delta f_{(2)}$ as defined in Eq.~(\ref{Full-DelF2}) in Eq.~(\ref{LandauFrame-Cond}), we obtain the following equation in terms of the undetermined variables, $\Delta u^\mu$, $\Delta T$ and $\Delta \mu$ as,
\begin{align}
    -\frac{I_{31}}{T} \Delta u^\mu + (I_{30}-\mu I_{20})\frac{\Delta T}{T^2} u^\mu + I_{20}\frac{\Delta \mu}{T} u^\mu &= A u^\mu + \mathcal{F}_1  \nabla^\mu \alpha + \mathcal{F}_2 \theta \nabla^\mu \alpha + \mathcal{F}_3 \sigma^{\mu}_\rho \dot{u}^\rho + \mathcal{F}_4 \sigma^\mu_\rho \nabla^\rho \alpha + \xi K_{31}\Delta^{\mu}_\rho \partial_{\gamma} \pi^{\rho\gamma} \nonumber\\
    &+ \mathcal{F}_5 D(\nabla^\mu \alpha) - \mathcal{F}_5 \omega^{\mu\rho} \nabla_\rho \alpha + \mathcal{F}_6 \nabla^\rho \sigma^\mu_\rho. \label{FirstMatch}
\end{align}
The various coefficients on the right-hand side of Eq.~(\ref{FirstMatch}) are expressed in terms of various thermodynamic integrals as follows\footnote{The term $A u^\mu$ cancels using Eq.~(\ref{secondMatch2}) and hence its form has no effect on our calculations. Hence, we have given the form of $A u^\mu$ in the appendix \ref{Appendix2} and not in the main body of the article for the sake of continuity.},
\begin{align}
    \mathcal{F}_1 &= ( K_{21}-\chi K_{31}), \label{B-Coef}\\
    \mathcal{F}_2 &= - K_{31} \frac{\partial (C\beta^2)}{\partial \beta} \frac{\beta}{3} + N_{31}\beta \frac{\chi}{3} -\frac{N_{21}\beta}{3}+ \frac{5}{3}\chi M_{42} - \frac{5}{3}M_{32} - \frac{L_{21}}{3} + L_{31}\frac{\beta}{3} \frac{\partial \chi}{\partial \beta} - \frac{1}{3}\mathcal{F}_5, \label{C-Coef}\\
    \mathcal{F}_3 &= 2\beta M_{42} - 4\beta L_{32} - 2\beta L_{31} - 2\beta^2 N_{32}, \label{D-Coef}\\
    \mathcal{F}_4 &= 2\beta A_{32} +2C\beta^3 K_{42} +2\chi M_{42} -2M_{32} - 2(\chi L_{42}-L_{32})\beta +2\chi L_{32}  +2 L_{22} - 2(\chi L_{42} - L_{32})\beta  \\
    &\,\,\,\,\,\,\,+ 2\chi \beta N_{32} - \mathcal{F}_5, \label{E-Coef}\\
    \mathcal{F}_5 &= \chi L_{31} - L_{21} - C\beta^2 K_{31}, \label{F-Coef}\\
    \mathcal{F}_6 &= 2\beta L_{32}. \label{H-Coef}
\end{align}
These coefficients were calculated under the assumption that $\tau_R(x,p)$ is a function of $\alpha$, $\beta$, and $u \cdot p$ without any specific parametrization of the relaxation time. This assumption necessitated the introduction of five new relaxation-time dependent thermodynamic integrals, $K_{nq}$, $L_{n,q}$, $M_{nq}$, $N_{nq}$ and $A_{nq}$, in comparison with the conventional RTA derivation. The explicit definitions of these integrals are discussed in Appendix~\ref{Appendix1}. In the RTA limit, these integrals can be expressed solely in terms of the relaxation time-independent thermodynamic integral, $I_{nq}$. Further, the second matching condition, $\epsilon=\epsilon_0$ imposes an additional constraint on the forms of $\Delta u^\mu$, $\Delta T$ and $\Delta \mu$ as,
\begin{align}
    u_\mu u_\nu \int \mathrm{dP} p^\mu p^\nu \Delta f_{(2)} = 0. \label{secondMatch}
\end{align}
This leads us to the following equation,
\begin{align}
        (I_{30} - \mu I_{20})\frac{\delta T}{T^2} + I_{20} \frac{\Delta \mu}{T} = \frac{I_{31}}{T} (u_\rho \Delta u^\rho) + A - \mathcal{F}_5 \dot{u}_\rho \nabla^\rho \alpha - \mathcal{F}_6 \sigma_{\rho\gamma}\sigma^{\rho\gamma}.\label{secondMatch2}
\end{align}
By Substituting Eq.~(\ref{secondMatch2}) into the first matching condition as defined in Eq.~(\ref{FirstMatch}), the variables $\Delta T$ and $\Delta \mu$ can be eliminated and the form of $\Delta u^\mu$ can be determined as,
\begin{align}
    \Delta u^\mu =& -\frac{T}{I_{31}}\Bigg[\bigg(\mathcal{F}_5 \dot{u}\cdot \nabla \alpha + \mathcal{F}_6\sigma:\sigma - \frac{I_{31}}{T}(u \cdot \Delta u)\bigg) u^\mu + \mathcal{F}_1 \nabla^\mu \alpha + \mathcal{F}_2 \theta \nabla^\mu \alpha + \mathcal{F}_3 \sigma^{\mu}_\rho \dot{u}^\rho + \mathcal{F}_4 \sigma^\mu_\rho \nabla^\rho \alpha + \xi K_{31}\Delta^{\mu}_\rho \partial_{\gamma} \pi^{\rho\gamma} \nonumber\\
    & + \mathcal{F}_5 D(\nabla^\mu \alpha) - \mathcal{F}_5 \omega^{\mu\rho} \nabla_\rho \alpha  + \mathcal{F}_6 \nabla^\rho \sigma^\mu_\rho\Bigg]. \label{Delta-u^m-Final}
\end{align}
Where we have used the notation, $\sigma : \sigma = \sigma^{\rho\gamma}\sigma_{\rho\gamma}$. It is important to emphasize that for a momentum-independent relaxation time, we obtain $\Delta u^\mu =0$, ensuring the consistency of the ERTA framework in the RTA limit.  Furthermore, we note that for a system with vanishing chemical potential ($\alpha \rightarrow 0$),  the form of $\Delta u^\mu$ reduces to the result presented in~\cite{Dash:2023ppc}. By Substituting Eq.~(\ref{Delta-u^m-Final}) in Eq.~(\ref{n^m*}), we obtain the $n^*_\mu$ contribution to the total number diffusion current as follows,
\begin{align}
   n_*^\mu =&  \frac{I_{21}}{I_{31}}\bigg[ (\mathcal{F}_5 \dot{u}\cdot \nabla \alpha + \mathcal{F}_6\sigma : \sigma)u^\mu + \mathcal{F}_1 \left(\nabla^\mu \alpha\right) + \mathcal{F}_2 \theta \left(\nabla^\mu \alpha\right) + \mathcal{F}_3 \sigma^{\mu}_\rho \dot{u}^\rho + \mathcal{F}_4 \sigma^\mu_\rho \nabla^\rho \alpha + \xi K_{31}\Delta^{\mu}_\rho \left(\partial_\gamma \pi^{\rho\gamma}\right) \nonumber\\
    & + \mathcal{F}_5 D(\nabla^\mu \alpha) -\mathcal{F}_5 \omega^{\mu\rho} \left(\nabla_\rho \alpha\right) + \mathcal{F}_6 \left(\nabla_\rho \sigma^{\mu\rho}\right) \bigg]. \label{n-star}
\end{align}
Similarly, by using Eq.~(\ref{delF-2ndOrder}) and Eq.~(\ref{Full-DelF2}) in Eq.~(\ref{form2}), we obtain the remaining contributions for the evolution of number diffusion as follows,
\begin{align}
    \bar{n}^\mu &= \left(\mathcal{J}_5 \dot{u}^\rho \nabla_\rho \alpha -2\beta L_{22}\sigma_{\rho\gamma}\sigma^{\rho\gamma}\right) u^\mu + \mathcal{J}_1 \nabla^\mu \alpha + \mathcal{J}_2 \theta \nabla^\mu \alpha + \mathcal{J}_3 \sigma^\mu_\rho \dot{u}^\rho + \mathcal{J}_4 \sigma^\mu_\rho \nabla^\rho \alpha - K_{21}\xi \Delta^\mu_\rho \partial_{\gamma} \pi^{\gamma\rho} + \mathcal{J}_5 D(\nabla^\mu \alpha) \nonumber\\
    &- \mathcal{J}_5 \omega^{\mu\rho} \nabla_\rho \alpha + \mathcal{J}_6 \nabla^\rho \sigma^\mu_\rho, \label{bar-n}
\end{align}
where the coefficients take the forms as follows,
\begin{align}
    \mathcal{J}_1 &= \chi K_{21} - K_{11} \label{Bn-Coef},\\
    \mathcal{J}_2 &= K_{21}\frac{\beta}{3} \frac{\partial (C\beta^2)}{\partial \beta} - \frac{\chi N_{21} \beta}{3} +\frac{N_{11}\beta}{3} - \frac{5}{3}\chi M_{32} + \frac{5}{3} M_{22} + \frac{L_{11}}{3} - L_{21}\frac{\beta}{3}\frac{\partial \chi}{\partial \beta} - \frac{\mathcal{J}_5}{3}, \label{Cn-Coef}\\
    \mathcal{J}_3 &= -2\beta M_{32} + 2\beta^2 N_{22} + 4\beta L_{22} + 2\beta L_{21},\label{Dn-Coef}\\
    \mathcal{J}_4 &= -2\beta(A_{22} + \chi N_{22}) -2C\beta^3 K_{32} - 2\chi M_{32} + 2M_{22} + 4\beta (\chi L_{32} - L_{22}) - 2\chi L_{22} -2L_{12} -\mathcal{J}_5, \label{En-Coef}\\
    \mathcal{J}_5 &= C\beta^2 K_{21} - (\chi L_{21}- L_{11}), \label{Fn-Coef}\\
    \mathcal{J}_6 &= -2\beta L_{22}. \label{Hn-Coef}
\end{align}
Employing Eq.~(\ref{n-star}) and  Eq.~(\ref{bar-n}) in  Eq.~(\ref{n^m-def}), we combine all the contributions and finally obtain the number diffusion evolution up to second order in gradients as\footnote{Here we should remark that the nature of Eq.~\eqref{n^m-evol} closely resembles that of the IReD approach described in Ref.~\cite{Wagner:2022ayd}.},
\begin{align}
    \dot{n}^{\langle \mu \rangle} +\frac{n^\mu}{\tau_n} &= \beta_n \nabla^\mu \alpha  - \lambda_{n\pi} \pi^{\mu\lambda}\nabla_\lambda \alpha - \tau_{n\pi} \pi^\mu_\lambda \dot{u}^\lambda - \delta_{nn} n^\mu \theta+l_{n\pi} \Delta^\mu_\rho \partial_{\gamma} \pi^{\rho\gamma}- \lambda_{nn} \sigma^\mu_\lambda n^\lambda - \lambda_\omega \omega^{\mu\lambda} n_\lambda. \label{n^m-evol}
\end{align}
The transport coefficients associated with the evolution of number diffusion are determined as,
\begin{align}
    \tau_n &= -\frac{\mathcal{G}_5}{\kappa}, \label{tauN}\\   
    \beta_n &= \frac{\mathcal{G}_1}{\tau_n}, \label{betaV}\\
    \lambda_{n\pi} &= -\frac{\mathcal{G}_6}{\tau_n} \left(\frac{\partial}{\partial \alpha} \left\{\frac{1}{2\eta}\right\}+ \chi \frac{\partial}{\partial \beta} \left\{\frac{1}{2\eta}\right\} \right), \label{lambdaVPi}\\
    \tau_{n\pi} &= -\frac{1}{\tau_n}\left(\frac{\mathcal{G}_3 + \mathcal{G}_6}{2\eta} - \mathcal{G}_6\beta \frac{\partial}{\partial \beta} \left\{\frac{1}{2\eta}\right\}\right), \label{tauVPi}\\
    \delta_{nn} &= -\frac{1}{\kappa\tau_n} \left(\mathcal{G}_2 + \mathcal{G}_5 \frac{\kappa \beta}{3} \frac{\partial}{\partial \beta} \left(\frac{1}{\kappa}\right)\right), \label{deltaVV}\\
    l_{n\pi} &= \frac{1}{\tau_n}\left(\frac{\mathcal{G}_6}{2\eta}-\xi K_{21}+\xi \frac{I_{21}}{I_{31}}K_{31}\right),  \label{lVPi}\\
    \lambda_{nn} &= -\frac{\mathcal{G}_4}{\kappa \tau_n}, \label{lambdaOm}\\
    \lambda_\omega &= \frac{\mathcal{G}_5}{\kappa \tau_n},
\end{align}
with $\mathcal{G}_i= \mathcal{J}_i + (I_{21}/I_{31})\mathcal{F}_i$ for  each of the above coefficients $i.e.$, $i=(1,2,3,4,5,6)$. The forms of $\mathcal{F}_i$ and $\mathcal{J}_i$ coefficients are defined in Eqs.~(\ref{B-Coef})-(\ref{H-Coef}) and Eqs.~(\ref{Bn-Coef})-(\ref{Hn-Coef}).

For the quantitative estimation of various momentum-dependent thermodynamic integrals, we employ the following parameterized form of $\tau_R(x,p)$:
\begin{align}
    &\tau_R(x,p) = \tau_0(x) \left(\frac{u\cdot p}{T}\right)^\ell,  &&\tau_0(x)=\frac{\bar{\kappa}}{T}, \label{tauX,P}
\end{align}
where we have introduced $\bar{\kappa}$, which  can be a function of $\mu/T$ depending on the type of interaction considered and $\tau_0(x)$ as a whole denotes the momentum-independent component of the relaxation time. Here, $\ell$ is the momentum dependence parameter that quantifies the order of momentum dependence of the relaxation time. The power-law parameterized form of $\tau_R(x,p)$ is motivated by the studies~\cite{Dusling:2009df, Chakraborty:2010fr, Dusling:2011fd, Kurkela:2017xis}. While the momentum dependence parameter $\ell$ in the relaxation time depends on the interaction dynamics within the system, its exact value for different physical scenarios remains to be determined. The range $0 \le \ell \le 1$ is seen to be relevant for the QCD medium~\cite{Dusling:2009df}. The region $\ell>1$ remains mostly unexplored except for the specific case of a gas of hard spheres \cite{Wagner:2023joq} which we have also discussed briefly in the following section.

\section{Results and discussion}\label{Section2}
\subsection{ERTA-modified particle diffusion relaxation
time and its relative significance}\label{sub1}
\begin{figure}
\begin{center}
\includegraphics[scale=0.55]{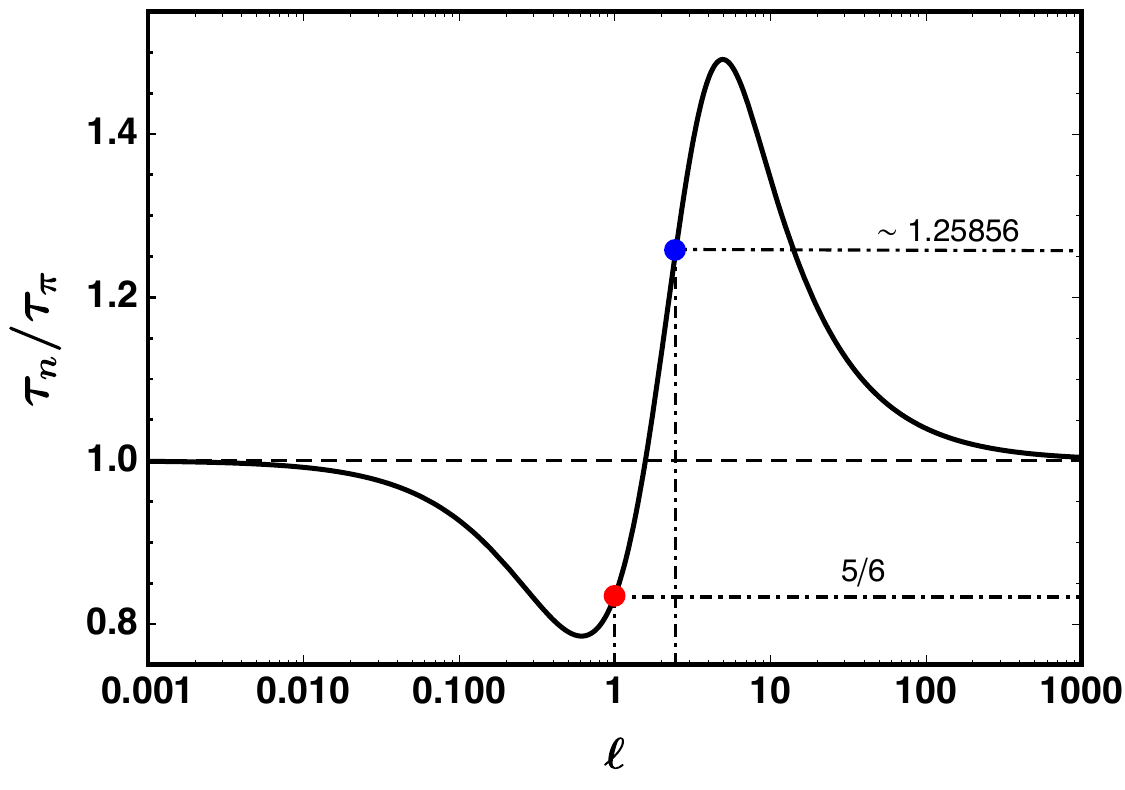}
\caption{  The ratio of the relaxation time for the diffusion mode to the relaxation time for the shear mode as a function of the momentum dependence parameter $\ell$. The red dot corresponds to the prediction for $\lambda \phi^4$ theory and the blue dot corresponds to this ratio for hard sphere scattering.}
\label{figure1}
\end{center}
\end{figure}
The relaxation time that sets the timescale for
the decay of the diffusion current is denoted by $\tau_n$. In the RTA limit $\ell \rightarrow 0$, we have $\tau_n \rightarrow \tau_0$. Our analysis shows that $\tau_n$ depends on the choice of the parameter $\ell$. In our previous study~\cite{Singh:2024leo}, we estimated the relaxation time for the evolution of the shear stress tensor, denoted as $\tau_\pi$, which determines the timescale for the decay of the shear viscous pressure, where $\tau_{\pi}$ is given by $\tau_{\pi}=L_{32}/K_{32}$. While the relaxation time for shear and number diffusion evolution in second-order is the same in the conventional RTA limit ($\ell=0$), we see that both the shear stress and number diffusion relaxation time are different from each other within the ERTA framework apart from the two limiting cases of $\ell \to 0, \infty$ and an intermediate value of $\ell \sim 1.57845 $. The transport coefficients derived in the previous section can be further simplified and can be explicitly expressed in terms of the momentum-dependence parameter $\ell$, once we fix the parametrization scheme, of which Eq.~\eqref{tauX,P} is only an example that has been widely used in the literature. Using Eq.~(\ref{tauN}) and the definitions of the thermodynamic integrals given in appendix \ref{Appendix1}, we obtain the relaxation time $\tau_n$ in terms of  $\ell$  as follows,
\begin{align}
    \tau_n &= -\frac{\Big[192 \Gamma(2(2+\ell))+16\Gamma^2(4+\ell) - 8\Gamma(4+\ell)\Gamma(5+\ell)+ \Gamma^2(5+\ell)-384\Gamma(3+2\ell)-24\Gamma(5+2\ell)\Big]\bar{\kappa}}{24T\Big[16\Gamma(3+\ell)-8\Gamma(4+\ell)+\Gamma(5+\ell)\Big]}, \quad \ell> -\frac{3}{2}.\label{tauN-Gamma}
\end{align}
Unlike in the RTA framework, we find that the ratio of shear relaxation time to the relaxation time for diffusion current deviates from unity and is dependent on the momentum-dependence parameter $\ell$ as,
\begin{align}
    \frac{\tau_n}{\tau_\pi} = \frac{(4+\ell)(3+\ell)}{24(\ell^2-\ell+4)}\left(\frac{48(2\ell^2-\ell+2)}{(4+2\ell)(3+2\ell)}-\frac{\ell^2 \Gamma^2(4+\ell)}{\Gamma(5+2\ell)} \right), \quad \quad l>-\frac{3}{2}.\label{relRatio}
\end{align}
As $\ell \to 0$, the ratio $\frac{\tau_n}{\tau_\pi}$ approaches unity, corresponding to the RTA limit where both the shear viscous pressure and diffusion current decays to their Navier-Stokes limits on the same timescale and a simple small $\ell$ expansion of this ratio behaves as,
\begin{align}
    &\ell\to0~{\rm limit:\qquad} \frac{\tau_n}{\tau_\pi} = 1 - \frac{5 \ell}{6} + \mathcal{O} \left(\ell^2\right).
\end{align}
However, this does not hold true for a finite value of $\ell$. For example, in $\lambda \phi^4$ theory (equivalent to $\ell = 1$ as showed in the next subsection), the ratio $\frac{\tau_n}{\tau_\pi}$ deviates from unity, taking the value $\frac{5}{6}$~\cite{Rocha:2023hts}. To understand the nature of relaxation times of the shear and diffusion processes of the given system, we have plotted the ratio $\frac{\tau_n}{\tau_\pi}$ for a wide range of values of $\ell$ in Fig.~\ref{figure1}. We notice that within the relevant region for QCD $i.e.$, $0\leq \ell \leq 1$, we have $\tau_n\leq \tau_\pi$, where the equality holds only at $\ell = 0$. This implies that the diffusion current reaches the Navier-Stokes limit quicker than the shear pressure for $0< \ell \leq 1$. However, this cannot be considered a general behavior of all fluids. The present work rather shows that the behavior of $\frac{\tau_n}{\tau_\pi}$ can vary wildly depending on the value of $\ell$.  Interestingly, this ratio reaches a global minimum value of $\frac{\tau_n}{\tau_\pi}\sim 0.785133$ at $\ell \sim 0.613843$, $i.e.$, within the region of interest for QCD matter. Going beyond this region, we note that the two relaxation times become equal at $\ell = 1.57845$.  For larger values of $\ell$, the ratio $\frac{\tau_n}{\tau_\pi}$ exceeds 1, reaching a maximum of approximately $1.49121$ at $\ell \sim 4.96012$ and then drops steadily towards unity asymptotically as $\ell\to \infty$. Another important observation from Fig.~\ref{figure1} is that the ratio $\frac{\tau_n}{\tau_\pi}$ appears to be bounded {(at least for $\ell\geq0$)} from both above and below.

As mentioned before, while the expected range of $\ell$ relevant to QCD matter is $0\leq\ell\leq1$, where we have $\tau_n\leq\tau_\pi$, there exists systems where we can have $\tau_n\geq \tau_\pi$. One such example is a gas of massless hard spheres, which was studied in Ref.~\cite{Wagner:2023joq}. Examining the information listed in Tables I and II of Ref.~\cite{Wagner:2023joq} we can find the ratio\footnote{We follow the IReD scheme to get this value.} $\frac{\tau_n}{\tau_\pi} \sim 1.25856$. This value of the ratio corresponds to the case, $\ell\sim2.47975$ in the present work. It should be noted though that the hard sphere system that we compared with was worked out in the ultrarelativistic limit and that massive hard sphere scattering will lead to different values of the above ratio. Thus, while the framework of ERTA can be used to describe the evolution of QCD-like matter, we may also use ERTA to describe fluids with a wider range of properties.

\subsection{Comparison of results with exact solution of \texorpdfstring{$\lambda \phi^4$}{~} theory and conventional RTA analysis}
{\renewcommand{\arraystretch}{2}%
\begin{table}
    \begin{tabular}{|c|c|c|c|c|}
     \hline   
      Coefficients  &ERTA ($\ell=1$)  & $\lambda \phi^4$ results & RTA ($\ell=0)$ \\
      \hline
      $\tau_n$  &  $\frac{20 d_g}{gn_0 \beta^2}$  &  $\frac{60}{g n_0 \beta^2}$ & $\tau_{0}$ \\
      \hline
      $\tau_n/\tau_\pi$  & $\frac{5}{6}$  & $\frac{5}{6}$ & $1$  \\
      \hline
      $\beta_n$  & $\frac{n_0}{20}$ & $\frac{n_0}{20}$ & $\frac{n_0}{12}$  \\
      \hline
      $2 \lambda_{n\pi}+ (\kappa/\eta) \lambda_{nn}$  & $\frac{19 \beta}{80}$ & $\frac{19\beta}{80}$ & $\frac{3\beta}{16}$ \\
     \hline    
     $\tau_{n\pi} + l_{n\pi}$  & $\frac{\beta}{20}$ & $\frac{\beta}{20}$ & $0$  \\
    \hline
    $l_{n\pi}$  & $\frac{\beta}{20}$ & $\frac{\beta}{40}$ & $0$  \\
    \hline
    $\delta_{nn}$  & $1$ & $1$ & $1$  \\
    \hline
    $\lambda_{\omega}$  & $-1$ & $-1$ & $-1$  \\
    \hline
\end{tabular}
\caption{Comparison of ERTA calculations with exact results from $\lambda \phi^4$ theory~\cite{Denicol:2022bsq,Rocha:2023hts} and standard RTA results ($\ell=0$). The value of $\tau_\pi$ was taken from the Ref.~\cite{Singh:2024leo}. The combinations of transport coefficients in the fourth and fifth rows are included to facilitate the comparison of coefficients with similar tensorial structures as explained in Appendix \ref{appendix3}.}
\label{t1}
\end{table}}
Exact calculations of transport coefficients by analytically solving the Boltzmann equation are very limited, as the Boltzmann equation is an integro-differential equation, making it significantly challenging to solve. The only known example in this regard, to the best of our knowledge, with an exact analytical solution, has been estimated only for an isotropic system with a relatively simple interaction strength~\cite{Denicol:2022bsq}. This model involves a system of massless scalar field particles with a $\lambda \phi^4$ tree-level self-interaction described by the Lagrangian as,
\begin{align}
    \mathcal{L} = \frac{1}{2} \partial^\mu \phi \partial_\mu \phi - \frac{1}{4!}\lambda \phi^4,
\end{align}
where $\lambda$ is the coupling constant of the scalar theory. The transition rate $W_{pp'\rightarrow kk'}$ can be estimated in this model and by employing Eq.~(\ref{collisionKernel}), the linearized collision term can be written as,
\begin{equation}
    C[f_k]=f_{0k}\hat{L} \phi_{\mathbf{k}}=f_{0k}\frac{g}{2} \int d K^{\prime} d P d P^{\prime} f_{0 \mathbf{k}^{\prime}}(2 \pi)^5 \delta^{(4)}\left(k+k^{\prime}-p-p^{\prime}\right)\left(\phi_{\mathbf{p}}+\phi_{\mathbf{p}^{\prime}}-\phi_{\mathbf{k}}-\phi_{\mathbf{k}^{\prime}}\right).
\end{equation}
Here, $\phi_k = (f_k -f_{0k})/f_{0k}$ and $g=\lambda^2 / (32\pi)$ is the coupling constant. The eigenfunctions and eigenvalues of the above collision operator $\hat{L}$ were determined in Refs.~\cite{Denicol:2022bsq,Rocha:2023hts} in terms of the associated  Laguerre polynomials $L_{n\textbf{k}}^{(2m+1)}$ of degree $n$. By expanding $\phi_k$ in terms of these eigenfunctions, the collision kernel can be written as,
\begin{equation}
C[f_k]=-\frac{u\cdot p}{\tau_R(x,p)}(f_k-f_{0k})+f_{0k}\frac{g \mathcal{M}}{2}\sum_{n, m=0}^{\infty} c_n^{\mu_1 \cdots \mu_{m}}\left(\frac{2}{n+m+1}-\delta_{m 0} \delta_{n 0}\right) L_{n \mathbf{k}}^{(2 m+1)} k_{\left\langle\mu_1\right.} \ldots k_{\left.\mu_{m}\right\rangle}, \label{phi4-collisionKernel}
\end{equation}
with $\mathcal{M}=e^\alpha/ (2\pi^2 \beta^2)$. The first term is the RTA term with a momentum-dependent relaxation time $\tau_R(x,p)$, which can be expressed as:
\begin{align}
    \tau_R(x,p)= \frac{4\pi^2}{g e^\alpha T}\left(\frac{u\cdot p}{T}\right).
\end{align}
It is evident that in Eq.~\eqref{phi4-collisionKernel} if we set $c_n^{\mu_1\cdots\mu_m} = 0$ for all values of $m$ and $n$, we get back the Anderson-Witting RTA, which does not lead to conservation laws when relaxation time is momentum dependent. However, one can ensure the conservation laws only if a certain number of the coefficients, $c_n^{\mu_1\cdots\mu_m}$ are kept as shown in Ref.~\cite{Denicol:2022bsq}. Using the method of moments it can be shown that, such counter terms, leading to novel RTA \cite{Rocha:2021zcw}, cancels the homogeneous part of $\delta f$. The quantity $\delta f^*$ in ERTA can be shown to do the same job of canceling the homogeneous part of $\delta f$, which allows the preservation of conservation laws.

From the above form of the relaxation time, we identify $\bar{\kappa}= 4\pi^2/(ge^\alpha)$ with $\alpha$ as the fugacity parameter and $\ell=1$, while mapping the scalar field theory to the current ERTA framework. This allows us to compare the ERTA transport coefficient estimations at $\ell=1$ with those calculated from the above exact results from the $\lambda \phi^4$ theory. This comparison of the transport coefficients associated with the number diffusion is summarized in Table \ref{t1}. The exact solutions of the transport coefficients were taken from Ref.~\cite{Rocha:2023hts}. The temperature behavior of all the coefficients derived from the ERTA framework is in good agreement with the exact results from the scalar theory. We observe that $\tau_n$ depends on the temperature and fugacity of the gas, for ERTA and  $\lambda \phi^4$ theory results. A similar dependence of the shear relaxation time, $\tau_\pi$ on the fugacity and temperature of the gas was also found in Ref.~\cite{Singh:2024leo}. We further note that, as demonstrated in the subsection~\ref{sub1}, the ratio $\frac{\tau_n}{\tau_\pi}$ at  $\ell=1$ exactly matches with the predictions from $\lambda \phi^4$ theory. We also note that the ERTA coefficient  $l_{n\pi}$ differ by a factor of $1/2$ while all other transport coefficients\footnote{Taking two combinations of the traditional transport coefficients as shown in Table \ref{t1} was necessary to compare our results with the $\lambda \phi^4$ results due to an ambiguity in the way these coefficients are defined in literature. A detailed discussion on why this particular choice was made has been outlined in Appendix \ref{appendix3}} associated with the number diffusion reduces to the predictions of $\lambda \phi^4$ theory in the limit ${\ell}=1$. Further, we also note that $\delta_{nn}$ and $\lambda_\omega$ takes the values $1$ and $-1$ respectively for all values of $\ell$.

To analyze the impact of the momentum-dependence of the relaxation time on the transport coefficients, we have compared the ERTA-modified transport coefficients (with a fixed value of $\ell = 1$) with the standard RTA results ($\ell = 0$).  The RTA results are presented in the fourth column of Table \ref{t1}. We observe that most coefficients associated with the number diffusion differ significantly with the inclusion of the momentum-dependent relaxation time. For instance,  the coefficient  $l_{n\pi}$  vanishes for a number-conserved massless system, as noted in Ref.~\cite{Jaiswal:2015mxa}. However, $l_{n\pi}$ is non-zero in the ERTA framework due to the particle energy dependence of the relaxation time, which aligns with the predictions of scalar field theory. 

\section{Summary and outlook}\label{Section3}
In conclusion, we have analyzed number diffusion in relativistic dissipative second-order hydrodynamics utilizing the ERTA framework for the first time. Using a Chapman-Enskog-like gradient expansion method at finite chemical potential, we solved the ERTA-modified Boltzmann equation. By incorporating the particle energy dependence of the collision time scale of the microscopic interactions through the relaxation time, we have derived the second-order evolution equation for the dissipative number diffusion current for a massless, particle number-conserving system. 

We have employed a power law parametrization to describe this
momentum dependence of the thermal relaxation time.  Our analysis highlights the sensitivity of the transport coefficients associated with the evolution of charge current to the momentum dependence of the relaxation time. Furthermore, we performed a comprehensive one-to-one comparison of the ERTA-derived transport coefficients with the exact results derived from a scalar field theory for a massless $\lambda \phi^4$ system with the appropriate choice of parametrization parameter for the relaxation time.

The present study underscores the critical importance of incorporating more realistic collision kernels into kinetic theory to improve the predictive power and accuracy of the hydrodynamic description of hot QCD nuclear matter from the underlying microscopic theory. This is particularly relevant in scenarios where net baryon charge diffusion plays a central role, such as in low-energy heavy-ion collisions. We also anticipate that the momentum dependence of the relaxation time might have a significant impact on the number diffusion in resistive second-order magneto-and spin-hydrodynamics. However, the formulation of the matching conditions and estimation of the non-equilibrium distribution function within the ERTA framework is challenging with a non-vanishing magnetic field and spin along with a finite chemical potential. We intend to explore these aspects in a future study. 

On the other hand, the fact that the present framework leads to different relaxation times for different transport processes naturally gives the opportunity to construct a theory of spin hydrodynamics from kinetic theory, where the spin degrees of freedom may relax at a rate different than other modes. This will be an important arena for exploration as primary studies indicate the spin relaxation time should be several times greater than other relevant relaxation timescales~\cite{Banerjee:2024xnd, Wagner:2024fry}. Such studies can shed light on the still unresolved issues of longitudinal spin polarization~\cite{STAR:2019erd}. Another important aspect of a system like QGP, dominated by strong interaction, is the implication of cross-diffusion among the multiple conserved charges such as electric, baryon, strangeness, etc. This area has received relatively less attention~\cite{Fotakis:2019nbq, Das:2021bkz, Dey:2024hhc}. Exploring the relaxation rates of these different diffusion processes will be an interesting area for future research.

\acknowledgments
{We sincerely thank Amaresh Jaiswal, Ashutosh Dash, Sunil Jaiswal, David Wagner, and M.S.A. Alam Khan for their valuable discussions and insights. S.B. acknowledges the financial support of the Faculty of Physics, Astronomy, and Applied Computer Science at Jagiellonian University, under Grant No. LM/36/BS.}

\appendix

\section{Thermodynamic Integrals and useful relations}\label{Appendix1}
In the present work, we define various thermodynamic integrals as follows:
\begin{align}
    I_{(r)}^{\mu_1\cdots\mu_n} &= \int \frac{dP}{\left(u\cdot p\right)^r} p^{\mu_1} \cdots p^{\mu_n} f_0, \label{I_r^mu1--mun} \\
    K_{(r)}^{\mu_1\cdots\mu_n} &= \int \frac{dP\, \tau_{\rm R}}{\left(u\cdot p\right)^r} p^{\mu_1} \cdots p^{\mu_n} f_0, \label{K_r^mu1--mun} \\
    L_{(r)}^{\mu_1\cdots\mu_n} &= \int \frac{dP\, \tau_{\rm R}^2}{\left(u\cdot p\right)^r} p^{\mu_1} \cdots p^{\mu_n} f_0, \label{L_r^mu1--mun} \\
    M_{(r)}^{\mu_1\cdots\mu_n} &= \int \frac{dP\, \tau_{\rm R}}{\left(u\cdot p\right)^r} \left(\frac{\partial\tau_{\rm R}}{\partial\left(u\cdot p\right)}\right) p^{\mu_1} \cdots p^{\mu_n} f_0, \label{M_r^mu1--mun} \\
    N_{(r)}^{\mu_1\cdots\mu_n} &= \int \frac{dP\, \tau_{\rm R}}{\left(u\cdot p\right)^r} \left(\frac{\partial\tau_{\rm R}}{\partial\beta}\right) p^{\mu_1} \cdots p^{\mu_n} f_0, \label{N_r^mu1--mun} \\
    A_{(r)}^{\mu_1\cdots\mu_n} &= \int \frac{dP\, \tau_{\rm R}}{\left(u\cdot p\right)^r} \left(\frac{\partial\tau_{\rm R}}{\partial\alpha}\right) p^{\mu_1} \cdots p^{\mu_n} f_0. \label{A_r^mu1--mun}
\end{align}
Each of the above momentum moments can be decomposed in terms of the hydrodynamic degrees of freedom as,
\begin{align}\label{Q-decom}
    Q^{\mu_1 \mu_2 ... \mu_n}_{(r)}= &Q^{(r)}_{n0} u^{\mu_1}u^{\mu_2}u^{\mu_3}...u^{\mu_n} + Q^{(r)}_{n1} (u^{\mu_1}u^{\mu_2}u^{\mu_3}....u^{\mu_{n-2}}\Delta^{\mu_{n-1}\mu_n} + \text{perm})  \nonumber\\
    &+ Q^{(r)}_{n2}(u^{\mu_1}u^{\mu_2}...u^{\mu_{n-4}}\Delta^{\mu_{n-3}\mu_{n-2}}\Delta^{\mu_{n-1}\mu_n} + \text{perm})+..... \nonumber\\
    &....+Q^{(r)}_{nq}(u^{\mu_1}\underbrace{\Delta^{\mu_{2}\mu_3}\Delta^{\mu_{4}\mu_5}....\Delta^{\mu_{n-1}\mu_n}}_{q -\Delta \; \text{ terms}}+\text{perm}),
\end{align}    
with $Q=\{I, K,L, M, N, A\}$ and $n\ge 2q$. Further, we define,
\begin{align}
    I_{nq}^{(r)} &= \frac{1}{\left(2q+1\right)!!} \int dP \left(u\cdot p\right)^{n-2q-r} \left(p\cdot\Delta\cdot p\right)^q f_0 \label{I_nq},\\
    K_{nq}^{(r)} &= \frac{1}{\left(2q+1\right)!!} \int dP\, \tau_{\rm R} \left(u\cdot p\right)^{n-2q-r} \left(p\cdot\Delta\cdot p\right)^q f_0 \label{K_nq},\\
    L_{nq}^{(r)} &= \frac{1}{\left(2q+1\right)!!} \int dP\, \tau_{\rm R}^2 \left(u\cdot p\right)^{n-2q-r} \left(p\cdot\Delta\cdot p\right)^q f_0 \label{L_nq},\\
    M_{nq}^{(r)} &= \frac{1}{\left(2q+1\right)!!} \int dP\, \tau_{\rm R} \left(\frac{\partial\tau_{\rm R}}{\partial\left(u\cdot p\right)}\right) \left(u\cdot p\right)^{n-2q-r} \left(p\cdot\Delta\cdot p\right)^q f_0 \label{M_nq},\\
    N_{nq}^{(r)} &= \frac{1}{\left(2q+1\right)!!} \int dP\, \tau_{\rm R} \left(\frac{\partial\tau_{\rm R}}{\partial\beta}\right) \left(u\cdot p\right)^{n-2q-r} \left(p\cdot\Delta\cdot p\right)^q f_0 \label{N_nq},\\
    A_{nq}^{(r)} &= \frac{1}{\left(2q+1\right)!!} \int dP\, \tau_{\rm R} \left(\frac{\partial\tau_{\rm R}}{\partial\alpha}\right) \left(u\cdot p\right)^{n-2q-r} \left(p\cdot\Delta\cdot p\right)^q f_0 \label{A_nq},
\end{align}
where $p\cdot \Delta \cdot p = \Delta^{\alpha\beta}p_\alpha p_\beta$. 
For any of the above integrals, we can omit the the superscript, $r$ when $r=0$ and express all of these integrals without $r$ by reducing $r$ to $0$ throughout the paper using the relation, $Q^{(r)}_{n,q} = Q^{(0)}_{n-r,q} \equiv Q_{n-r,q}$. Another useful relation for the current analysis can be derived using:
\begin{align}
    \tau_R \frac{\partial \tau_R}{\partial \beta} & = \tau_R \frac{\partial}{\partial \beta} \left[\tau_0(x) \left(\frac{u\cdot p}{T}\right)^\ell \right]
    = \tau_R^2 \left[\left(\frac{\partial \ln\tau_0}{\partial \beta}\right) + T \ell\right], 
\end{align}
where $\tau_R (x,p)$ takes the form as in Eq.~(\ref{tauX,P}). 
This leads to the following relation between $N_{nq}$ and $L_{nq}$ integrals as:
\begin{align}
    N_{nq}
    = \left(\frac{\partial \ln\tau_0}{\partial \beta}\right) L_{nq} + T\ell L_{nq}.
\end{align}
For $\tau_0(x)=$ constant, we have $N_{nq}=0$ when $\ell\rightarrow 0$ and for $\tau_0(x) = \bar{\kappa}/T$, we obtain $N_{nq} = T(\ell+1)L_{nq}$ for $\bar{\kappa} = $ constant.  Similarly, we employ the relation:
\begin{align}
    \tau_R\frac{\partial \tau_R}{\partial \alpha} &
    = \tau_R^2 \left(\frac{\partial \ln\tau_0}{\partial \alpha}\right)
    ,\label{Ap1}
\end{align}
By employing Eq.~(\ref{Ap1}) in  Eq.~(\ref{A_nq}), we can relate $A_{nq}$ with the $L_{nq}$ integral as,
\begin{align}
    A_{nq} 
    = \left(\frac{\partial \ln\tau_0}{\partial \alpha}\right) L_{nq}.
\end{align}
The above expression is useful when comparing the ERTA results with the $\lambda \phi^4$ theory where $\tau_0 (x)$ is a function of both temperature and the fugacity of the gas,
\begin{align}
    \tau_0(x) = \frac{4\pi^2}{g e^\alpha T}.
\end{align}
When $\tau_R \rightarrow \tau_c$ in the case of RTA, we see that the above integral $A_{nq}$ vanishes. We also have the following relation for any of these integrals in the massless limit:
\begin{align}
    Q^{(r)}_{n,q}=-\left(\frac{1}{2q+1}\right)Q^{(r)}_{n,q-1}.
\end{align}

\section{Definition of A in Eq.~(\ref{FirstMatch})}\label{Appendix2}

The term $A u^\mu$ appearing in Eq.~(\ref{FirstMatch}) can be written as,
\begin{align}
    A u^\mu= \Bigg[(I_{30} - I_{31})\frac{(u\cdot \delta u)}{T} - &(K_{30} \dot{\beta} - K_{20} \dot{\alpha} + K_{31}\beta \theta) + \bar{A}_\sigma \sigma^2 + \bar{A}_{\alpha 1} (\nabla\alpha)^2 + \bar{A}_{\alpha 2} \nabla^2 \alpha  \nonumber\\
    &+ \bar{A}_{u\alpha} \dot{u}\cdot \nabla\alpha + L_{31} \nabla \chi \cdot \nabla \alpha - K_{31} \nabla\alpha \cdot \nabla (C\beta^2)\Bigg] u^\mu
\end{align}
Where,
\begin{align}
    \bar{A}_\sigma &= 2(\beta M_{42} - \beta L_{31} + \beta L_{32} - \beta L_{42} - \beta L_{32}) \nonumber\\
    \bar{A}_{\alpha 1} &= \chi^2 N_{31} - \chi N_{21} -C\beta^2 K_{31} + \chi C \beta^2 K_{41} + \chi A_{31} - A_{21} - \chi^2 L_{41} + 2 \chi L_{31} - L_{21} -\frac{C^2 \beta^4 I_{41}}{2}\nonumber\\
    \bar{A}_{\alpha 2} &= -C\beta^2 K_{31} - L_{21} + \chi L_{31}\nonumber\\
    \bar{A}_{u \alpha} &= -\chi \beta N_{31} + \beta N_{21} + \chi M_{41} - M_{31}- \chi L_{30} + \chi L_{31} + L_{20} {\color{red}{-}} C\beta^2 (K_{31}-K_{30})
\end{align}
The above form for $A$ is given for the sake of completeness since it cancels out in our calculations following Eq.~(\ref{secondMatch2}).

\section{Combination of coefficients in Table \ref{t1} and related ambiguity} \label{appendix3}

Using the relation,
\begin{align}\label{pi-derivative}
    \Delta^\mu_\rho \partial_\gamma \pi^{\rho \gamma} &= - \pi^{\mu\gamma}\dot{u}_\gamma + \Delta^\mu_\rho \nabla_\gamma \pi^{\rho\gamma},
\end{align}
we can re-write the evolution equation for the number diffusion current given by Eq.~(\ref{n^m-evol}) as,
\begin{align}\label{mod-n^m-evo}
    \dot{n}^{\langle \mu \rangle} + \frac{n^\mu}{\tau_n} &= \beta_n \nabla^\mu \alpha - \left(2\lambda_{n\pi} + \frac{\kappa}{\eta} \lambda_{nn}\right) \pi^\mu_\lambda \nabla^\lambda \alpha - \left(\tau_{n\pi}+ l_{n\pi}\right) \pi^{\mu\lambda}\dot{u}_\lambda -\delta_{nn}n^\mu \theta + l_{n\pi} \Delta^\mu_\rho \nabla_\gamma \pi^{\rho \gamma} -\lambda_\omega \omega^{\mu\lambda}n_\lambda
\end{align}
where we have used Eq.~(\ref{eta}) and Eq.~(\ref{kappa}) along with Eq.~(\ref{pi-derivative}) to simplify the number diffusion evolution into its independent tensorial structures. As we can see, since the combination $\tau_{n\pi} + l_{n\pi}$ is the coefficient of the tensor $\pi^{\mu\lambda}\dot{u}_\lambda$, there exists an ambiguity in how both $\tau_{n\pi}$ and $l_{n\pi}$ are separately defined. A transformation of both of these coefficients by $\tau_{n\pi} \rightarrow \tau_{n\pi}+ \Phi$ and $l_{n\pi} \rightarrow l_{n\pi}-\Phi$ will not change the evolution equation Eq.~(\ref{mod-n^m-evo}), but it will change these coefficients separately. Hence, in order to avoid this ambiguity while comparing with previous calculations in literature, we have decided to take the combinations of these coefficients belonging to independent tensorial structures in the comparision Table \ref{t1}. A similar logic follows for the combination of co-efficients of $\pi^\mu_\lambda \nabla^\lambda \alpha$. 

\bibliography{ref}
\end{document}